\documentclass[runningheads]{llncs}

\usepackage[printwatermark]{xwatermark}
\usepackage{graphicx}
\usepackage{todonotes}
\usepackage{booktabs}
\usepackage{tikz, float}
\usepackage{paralist}
\usepackage{subcaption}
\usepackage{tikz}
\usetikzlibrary{trees, arrows}
\usepackage{comment}
\usepackage{listings}
\usepackage[ruled]{algorithm2e}
\usepackage{amsmath}

\newcommand{\ReinitPP}{Reinit$^{++}$\xspace}

\begin{document}

\title{\ReinitPP: Evaluating the Performance of Global-Restart Recovery
Methods For MPI Fault Tolerance}

\titlerunning{\ReinitPP: Global-Restart Recovery For MPI Fault Tolerance}

\author{
    Giorgis Georgakoudis\inst{1} \and
    Luanzheng Guo\inst{2}\thanks{Work performed during internship
    at Lawrence Livermore National Laboratory}
    \and
    Ignacio Laguna\inst{1}
}

\authorrunning{G. Georgakoudis et al.}

\institute{
    \textit{Center for Advanced Scientific Computing},\\
    Lawrence Livermore National Laboratory, USA,\\
    \email{\{georgakoudis1, lagunaperalt1\}@llnl.gov}
    \and
    \textit{EECS}, UC Merced, USA,\\
    \email{lguo4@ucmerced.edu}
}

\maketitle

\begin{abstract}
    Scaling supercomputers comes with an increase in failure rates due to
the increasing number of hardware components. In standard practice,
applications are made resilient through checkpointing data and
restarting execution after a failure occurs to resume from the latest
checkpoint. However, re-deploying an application incurs overhead by
tearing down and re-instating execution, and possibly limiting
checkpointing retrieval from slow permanent storage.

In this paper we present \ReinitPP, a new design and implementation of
the Reinit approach for global-restart recovery, which avoids
application re-deployment. We extensively evaluate
\ReinitPP contrasted with the leading MPI fault-tolerance approach of
ULFM, implementing global-restart recovery, and the typical practice of
restarting an application to derive new insight on performance.
Experimentation with three different HPC proxy applications made
resilient to withstand process and node failures shows that \ReinitPP
recovers much faster than restarting, up to 6$\times$, or ULFM, up to
3$\times$, and that it scales excellently as the number of MPI processes
grows.

\end{abstract}

\section{Introduction}

HPC system performance scales by increasing the number of computing
nodes and by increasing the processing and memory elements of each node.
Furthermore, electronics continue to shrink, thus are more susceptible
to interference, such as radiation upsets or voltage fluctuations.
Those trends increase the probability of a failure happening, either due
to component failure or due to transient soft errors affecting
electronics.  Large HPC applications run for hours or days and use most,
if not all, the nodes of a supercomputer, thus are vulnerable to
failures, often leading to process or node crashes.  Reportedly, the
mean time between a node failure on petascale systems has been measured
to be 6.7 hours~\cite{6903615}, while worst-case
projections~\cite{Dongarra:2011:IES:1943326.1943339} foresee that
exascale systems may experience a failure even more frequently.

HPC applications often implement fault tolerance using checkpoints
to restart execution, a method referred to as \emph{Checkpoint-Restart}
(CR).  Applications periodically store checkpoints, e.g., every few
iterations of an iterative computation, and when a failure occurs,
execution aborts and restarts again to resume from the latest
checkpoint.  Most scalable HPC applications follow the Bulk Synchronous Parallel
(BSP) paradigm, hence CR with global, backward, non-shrinking
recovery~\cite{Laguna:2014}, also known as \emph{global-restart} 
naturally fits their execution. CR is straightforward to implement but
requires re-deploying the whole application on a failure, re-spawning all
processes on every node and re-initializing any application data
structures. This method has significant overhead since a failure of few
processes, even a single process failure, requires complete
re-deployment, although most of the processes survived the failure.

By contrast,  User-level Fault Mitigation (ULFM)~\cite{bland2013post}
extends MPI with interfaces for handling failures at the application
level without restarting execution. The programmer is required to use
the ULFM extensions to detect a failure and repair communicators and
either spawn new processes, for non-shrinking recovery, or continue
execution with any survivor processes, for shrinking recovery. Although
ULFM grants the programmer great flexibility to handle failures, it
requires considerable effort to refactor the application for 
correctly and efficiently implementing recovery.

Alternatively, Reinit~\cite{laguna2016evaluating,doi:10.1002/cpe.4863}
has been proposed as an easier-to-program approach, but equally capable
of supporting global-restart recovery.  Reinit extends MPI with a
function call that sets a rollback point in the application. It
transparently implements MPI recovery, by spawning new processes and
mending the world communicator at the MPI runtime level. Thus, Reinit
transparently ensures a consistent, initial MPI state akin to the state
after MPI initialization. However, the existing implementation of
Reinit~\cite{doi:10.1002/cpe.4863} is hard to deploy, since it requires
modifications to the job scheduler, and difficult to compare with ULFM,
which only requires extensions to the MPI library. Notably, both Reinit
and ULFM approaches assume the application has checkpointing in place to resume
execution at the application level.

Although there has been a large
bibliography~\cite{bland2013post,doi:10.1002/cpe.4863,laguna2016evaluating,bland2015lessons,losada2017resilient,pauli2014fault,hori2015sliding,katti2015scalable,herault2015practical,bouteiller2015plan,Laguna:2014:EUF:2642769.2642775}
discussing the programming model and prototypes of those approaches,
no study has presented an in-depth performance evaluation of them
--most 
previous works either focus on individual aspects of each approach or perform 
limited scale experiments. 
In this paper, we present an extensive evaluation using HPC proxy
applications to contrast these two leading global-restart recovery approaches.
Specifically, our contributions are:
\begin{itemize}
  \item A new design and implementation of the Reinit approach, named
    \ReinitPP, using the latest Open MPI runtime. Our design and
    implementation supports recovery from either process or node
    failures, is high performance, and deploys easily by extending the
    Open MPI library.  Notably, we present a precise definition of the
    failures it handles and the scope of this design and implementation.
  \item An extensive  evaluation of the performance of the possible recovery
    approaches (CR, \ReinitPP, ULFM) using three HPC proxy
    applications (CoMD, LULESH, HPCCG), and including file and in-memory
    checkpointing schemes.
  \item New insight from the results of our evaluation which show that
    recovery under \ReinitPP is up to 6$\times$ faster than CR and up to
    3$\times$ faster than ULFM. Compared to CR, \ReinitPP avoids the
    re-deployment overhead, while compared to UFLM, \ReinitPP avoids
    interference during fault-free application execution and has less
    recovery overhead.
\end{itemize}


\section{Overview}
\label{sec:overview}

This section presents an overview of the state-of-the-art approaches for
MPI fault tolerance. Specifically, it provides an overview of the
recovery models for applications and briefly discusses ULFM and Reinit,
which represent the state-of-the-art in MPI fault tolerance.

\subsection{Recovery Models for MPI Applications}

There are several models for fault tolerance depending on the
requirements of the application.  Specifically, if all MPI processes
must recover after a failure, recovery is \emph{global}; otherwise if
some, but not all, of the MPI processes need to recover then recovery is
deemed as \emph{local}.  Furthermore, applications can either recover by
rolling back computation at an earlier point in time, defined as
\emph{backward} recovery, or, if they can continue computation 
without backtracking, recovery is deemed as \emph{forward}. Moreover, if
recovery restores the number of MPI processes to resume execution, it is
defined as \emph{non-shrinking}, whereas if execution continues with
whatever number of processes surviving the failure, then recovery is
characterized as \emph{shrinking}.
\emph{Global-restart} implements global, backward, non-shrinking
recovery which fits most HPC applications that follow a bulk-synchronous
paradigm where MPI processes have interlocked dependencies, thus it is
the focus of this work.

\subsection{Existing Approaches for MPI Fault Tolerance}

\subsubsection{ULFM}
One of the state-of-the-art approaches for fault tolerance in MPI is User-level
Fault Mitigation (ULFM)~\cite{bland2013post}. ULFM extends MPI to enable
failure detection at the application level and provide a set of
primitives for handling recovery. Specifically, ULFM taps to the
existing error handling interface of MPI to implement user-level
fault notification. Regarding its extensions to the MPI interface, we elaborate on communicators
since their extensions are a superset of other
communication objects (windows, I/O).
Following, ULFM extends MPI with a \emph{revoke} operation
(\verb|MPI_Comm_revoke(comm)|) to invalidate a communicator such that
any subsequent operation on it raises an error.  Also, it defines a
\emph{shrink} operation (\verb|MPI_Comm_shrink(comm, newcomm)|) that
creates a new communicator from an existing one after excluding any
failed processes. Additionally, ULFM defines a collective \emph{agreement}
operation (\verb|MPI_Comm_agree(comm,flag)|) which achieves consensus on
the group of failed processes in a communicator and on the value of the
integer variable \verb|flag|.

Based on those extensions, MPI programmers are
expected to implement their own recovery strategy tailored to their
applications. ULFM operations are general enough to implement any type of
recovery discussed earlier.
However, this generality comes at the cost of complexity. Programmers
need to understand the intricate semantics of those operations to
correctly and efficiently implement recovery and restructure, possibly
significantly, the application for explicitly handling
failures.
Although ULFM provides examples that prescribe the implementation of
global-restart, the programmer must embed this in the code and refactor
the application to function with the expectation that communicators may
change during execution due to shrinking and merging, which is not
ideal.

\subsubsection{Reinit}

Reinit~\cite{doi:10.1177/1094342015623623,doi:10.1002/cpe.4863} has
been proposed as an alternative approach for implementing global-restart
recovery, through a simpler interface compared to ULFM.
The most recent implementation~\cite{doi:10.1002/cpe.4863}
of Reinit is limited in several aspects: 
\begin{inparaenum}[(1)]
\item it requires modifying the job scheduler (SLURM), besides the MPI
  runtime, thus it is impractical to deploy and skews performance
  measurements due to crossing the interface between the job scheduler
  and the MPI runtime; 
\item its implementation is not publicly available;
\item it bases on the MVAPICH2 MPI runtime, which makes comparisons with
  ULFM hard, since ULFM is implemented on the Open MPI runtime.
\end{inparaenum}
Thus, we opt for a new design and implementation\footnote{Available
open-source at \url{https://github.com/ggeorgakoudis/ompi/tree/reinit}},
named \ReinitPP, which we present in detail in the next section.

\section{\ReinitPP}
\label{sec:method}

This section describes the programming interface of \ReinitPP, the
assumptions for application deployment, process and node failure
detection, and the recovery algorithm for global-restart. We also
define the semantics of MPI recovery for the implementation of 
\ReinitPP as well as discuss its specifics.

\subsection{Design}

\begin{figure}[t]
  \centering
  \resizebox{0.85\textwidth}{!}{
  \input{fig-api.tex}
}
  \caption{The programming interface of \ReinitPP}
  \label{fig:api}
\end{figure}

\begin{figure}[t]
  \centering
  \resizebox{0.85\textwidth}{!}{
  \input{fig-sample.tex}
}
  \caption{Sample usage of the interface of \ReinitPP}
  \label{fig:sample}
\end{figure}

\subsubsection{Programming Interface of \ReinitPP}
Figure~\ref{fig:api} presents the programming interface of \ReinitPP in the
C language, while figure~\ref{fig:sample} shows sample usage of it. There
is a single function call, \verb|MPI_Reinit|, for the programmer to call
to define the point in code to rollback and resume execution after a
failure. This function  must be called after \verb|MPI_Init| so ensure
the MPI runtime has been initialized.  Its arguments imitate the
parameters of \verb|MPI_Init|, adding a parameter for a pointer to a
user-defined function.  \ReinitPP expects the programmer to encapsulate in
this function the main computational loop of the application, which is
restartable through checkpointing.  Internally, \verb|MPI_Reinit| passes
the parameters \verb|argc| and \verb|argv| to this user-defined
function, plus the parameter \verb|state|, which indicates the MPI state
of the process as values from the enumeration type
\verb|MPI_Reinit_state_t|.  Specifically, the value
\verb|MPI_REINIT_NEW| designates a new process executing for the first
time, the value \verb|MPI_REINIT_REINITED| designates a survivor process
that has entered the user-defined function after rolling back due to a
failure, and the value \verb|MPI_REINIT_RESTARTED| designates that the
process has failed and has been re-spawned to resume execution.  Note
that this state variable describes only the MPI state of \ReinitPP, thus
has no semantics on the application state, such as whether to load a
checkpoint or not. 


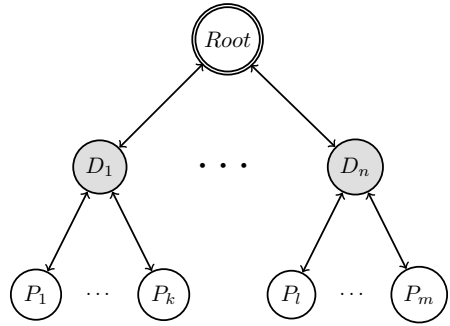
\begin{figure}[t]
  \centering
  \resizebox{0.5\textwidth}{!}{
  \begin{tikzpicture}
  [
      grow                    = down,
      <->, thick,
      treenode/.style = {align=center, text centered, circle, draw, font=\normalsize},
      level/.style = {sibling distance = 4cm/#1, level distance = 2cm},
      sloped
    ]
    \node [treenode, double] {$Root$}
    child { node [treenode, fill=gray!25] (D1) {$D_1$} 
      child { node [treenode] (P11) {$P_{1}$} }
      child { node [treenode] (P1n) {$P_{k}$} }
    }
    child { node [treenode, fill=gray!25] (Dn) {$D_n$}
      child { node [treenode] (Pn1) {$P_{l}$} }
      child { node [treenode] (Pnn) {$P_{m}$} }
    };

    \path (D1) -- (Dn) node [midway] {\huge $\cdots$};
    \path (P11) -- (P1n) node [midway] {$\cdots$};
    \path (Pn1) -- (Pnn) node [midway] {$\cdots$};
\end{tikzpicture}
}
  \caption{Application deployment model}
  \label{fig:deployment}
\end{figure}

\subsubsection{Application Deployment Model}
\ReinitPP assumes a logical, hierarchical topology of application
deployment. 
Figure~\ref{fig:deployment} shows a graphical representation
of this deployment model. At the top level, there is a single \emph{root}
process that spawns and monitors \emph{daemon} processes, one on each of
the computing nodes reserved for the application. Daemons spawn and
monitor \emph{MPI processes} local to their nodes. 
The root communicates with daemons and keeps track
of their liveness, while daemons track the liveness of their children 
MPI processes. Based on this execution and deployment model, \ReinitPP
performs fault detection, which we discuss next.

\subsubsection{Fault Detection}
\ReinitPP targets \emph{fail-stop} failures of either MPI processes or
daemons. A daemon failure is deemed equivalent to a node failure. The causes for
those failures may be transient faults or hard faults of hardware
components.

In the design of \ReinitPP, the root manages the execution of the whole
applications, so any recovery decisions are taken by it, hence it
is the focal point for fault detection. Specifically, if an MPI process
fails, its managing daemon is notified of the failure
and forwards this notification to the root, without taking an action
itself. If a daemon process fails, which means either the node failed or
the daemon process itself, the root directly detects the failure and
also assumes that the children MPI processes of that daemon are lost too.
After detecting a fault the root process proceeds with recovery, which
we introduce in the following section.

\subsubsection{MPI Recovery}

\ReinitPP recovery for both MPI process and daemon failures is similar,
except that on a daemon failure the root chooses a new host node to
re-instate failed MPI processes, since a daemon failure proxies a node
failure. For recovery, the root process broadcasts a \emph{reinit} message
to all daemons. Daemons receiving that message roll back survivor
processes and re-spawn failed ones. After rolling back survivor MPI processes 
and spawning new ones, the semantics of MPI recovery are that only the
world
communicator is valid and any previous MPI state (other communicators,
windows, etc.) has been discarded.
This is similar to the MPI state available immediately after an
application calls \verb|MPI_Init|.
Next, the application restores its
state, discussed in the following section.

\subsubsection{Application Recovery}
\ReinitPP assumes that applications are responsible for saving and
restoring their state to resume execution.  Hence, both survivor and
re-spawned MPI processes should load a valid checkpoint after MPI
recovery to restore application state and resume computation.


\subsection{Implementation}

We implement \ReinitPP in the latest Open MPI runtime, version 4.0.0. The
implementation supports recovery from both process and daemon (node)
failures. This implementation does not presuppose any particular job
scheduler, so it is compatible with any job scheduler the Open MPI runtime
works with. Introducing briefly the Open MPI software architecture, 
it comprises of three frameworks of distinct functionality:
\begin{inparaenum}[(i)]
\item the OpenMPI MPI layer (OMPI), which implements the interface of
  the MPI specification used by the application developers;
\item the OpenMPI Runtime Environment (ORTE), which implements runtime
  functions for application deployment, execution monitoring, and fault 
  detection, and
\item the Open Portability Access Layers (OPAL), which implements
  abstractions of OS interfaces, such as signal handling, process
  creation, etc.
\end{inparaenum}

\ReinitPP extends OMPI to provide the function
\verb|MPI_Reinit|. It extends ORTE to propagate fault notifications from
daemons to the root and to implement the mechanism of MPI recovery on detecting a fault.
Also, \ReinitPP extends OPAL to implement low-level process signaling for
notifying survivor process to roll back.  The following sections provide
more details.

\subsubsection{Application Deployment}
\ReinitPP requires the application to deploy using the default launcher of
Open MPI, \verb|mpirun|. Note that using the launcher \verb|mpirun| is
compatible with any job scheduler and even
uses optimized deployment interfaces, if the scheduler provides any.
Physical application deployment in Open MPI closely follows the logical
model of the
design of \ReinitPP. Specifically, Open MPI sets the root of the deployment at the
process launching the \verb|mpirun|, typically on a login node of
HPC installations, which is deemed as the Head Node Process (HNP) in Open
MPI terminology. Following, the root launches an ORTE daemon on each node 
allocated for the application.
Daemons spawn the set of MPI processes in each node and 
monitor their execution. The root process
communicates with each daemon over a channel of a reliable network transport and
monitors the liveness of daemons through the existence of this channel.

Launching an application, the user specifies the number of MPI processes
and optionally the number of nodes (or number of processes per node).
To withstand process failures, this specification of deployment is
sufficient, since \ReinitPP re-spawns failed processes on their original
node of deployment. However, for node failures, the user must
\emph{over-provision} the allocated process slots for re-spawning the set
of MPI processes lost due to a failed node. To do so, the most
straightforward way is to allocate more nodes than required for fault-free
operation, up to the maximum number of node failures 
to withstand. 

\subsubsection{Fault Detection}
In Open MPI, a daemon is the parent of the MPI processes on its node. If
an MPI process crashes, its parent daemon is notified, by trapping the
signal \verb|SIGCHLD|, in POSIX semantics. Implementing the fault
detection requirements of \ReinitPP, a daemon relays the fault notification
to the root process for taking action.
Regarding node failures, the root directly detects them proxied through
daemon failures. Specifically, the root has an open communication channel with each daemon over
some reliable transport, e.g., TCP. If the connection over that
communication channel breaks, the root process is notified of the
failure and regards the daemon at fault, thus assuming all its children
MPI process lost and its host node is unavailable.
For both types of failures (process and node), the root process
initiates MPI recovery.


\begin{algorithm}[t]
  \DontPrintSemicolon
  \caption{Root: HandleFailure\label{alg:df}}
  \label{algo:root}
  \KwData{
    $\mathcal{D} \colon$ the set of daemons, \\
    $Children(x) \colon $ returns the set of children MPI processes of daemon $x$, \\
    $Parent(x) \colon$ returns the parent daemon of MPI process $x$
  }
  \KwIn{The failed process $f$ (MPI process or daemon)}
  \tcp{failed process is a daemon}
  \uIf{$f \in \mathcal{D}$}{
    $\mathcal{D} \leftarrow \mathcal{D} \setminus \{f\}$\;
    $d' \leftarrow d \mid \underset{d \in \mathcal{D}}{\arg\min}~Children(d) $\;
    \tcp{broadcast REINIT to all daemons}
    Broadcast $\mathcal{D}$ message $\big\langle \texttt{REINIT}, ~\left\{~\langle d', c \rangle ~|~ \forall c \in Children(f)~\right\} \big\rangle$
  }
  \tcp{failed process is an MPI process}
  \Else {
    Broadcast $\mathcal{D}$ message $\big\langle \texttt{REINIT}, ~\left\{~\langle Parent(f), f \rangle ~\right\} \big\rangle$
  }
\end{algorithm}

\begin{algorithm}[t]
  \DontPrintSemicolon
  \caption{Daemon $\hat{d}$: HandleReinit}
  \KwData{$Children(x) \colon $ returns the set of children MPI processes of daemon $x$, \\
  $Parent(x) \colon$ returns the parent daemon of MPI process $x$}
  \label{algo:daemon}
  \KwIn{ List $\{ \langle d_i, c_i \rangle, \cdots \}$ }
  \tcp{Signal survivor MPI processes}
  \For{$c \in Children(\hat{d})$}{
    $c.state \leftarrow$ \texttt{MPI\_REINIT\_REINITED}\;
    Signal \texttt{SIGREINIT} to $c$\;
  }
  \tcp{Spawn new process if $\hat{d}$ is parent}
  \ForEach{ $ \{ \langle d_i, c_i \rangle, \cdots\} $ } {
    \If{ $\hat{d} == d_i$ }{
      $Children(\hat{d}) \leftarrow Children(\hat{d}) \cup c_i$\;
      $c_i.state \leftarrow$ \texttt{MPI\_REINIT\_RESTARTED}\;
      Spawn $c_i$\;
    }
  }
\end{algorithm}

\begin{algorithm}[t]
  \DontPrintSemicolon
  \caption{\ReinitPP internals}
  \label{algo:mpi}
  \SetKwFunction{FReinit}{MPI\_Reinit}
  \SetKwFunction{FFoo}{foo}
  \SetKwFunction{FSig}{OnSignalReinit}
  \SetKwProg{Fn}{Function}{:}{end}
  \SetKw{Goto}{goto}

  \Fn{\FSig{}}{
    \Goto Rollback\;
  }

  \Fn{\FReinit{$argc,argv,$ \FFoo}}{
    Install signal handler \FSig on \texttt{SIGREINIT}\;
    \nlset{Rollback:}
    \label{RollbackPoint}
    \If{ $this.state == $ \texttt{MPI\_REINIT\_REINITED} }{
      Discard MPI state\;
      Wait on barrier\;
      Re-initialize world communicator\;
    }

    \KwRet{ \FFoo($argc,argv,this.state$) }\;
  }

\end{algorithm}

\subsubsection{MPI Recovery}
Algorithm~\ref{algo:root} shows in pseudocode the operation of the root
process when handling a failure.  On detecting a failure, the root
process distinguishes whether it is a faulty daemon or MPI process.  For
a node failure, the root selects the \emph{least loaded node} in the
resource allocation, that is the node with the fewest occupied process
slots, and sets this node's daemon as the parent daemon for failed
processes.  For a process failure, the root selects the original parent
daemon of the failed process to re-spawn that process.  Next, the root process
initiates recovery by broadcasting to all daemons a message with the
\verb|REINIT| command and the list of processes to spawn, along with
their selected parent daemons.
Following, when a daemon receives that message
it signals its survivor, children MPI processes
to roll back, and re-spawns any processes in the list that have
this daemon as their parent. Algorithm~\ref{algo:daemon} presents this
procedure in pseudocode.

Regarding the asynchronous, signaling interface of \ReinitPP, Algorithm~\ref{algo:mpi} illustrates 
the internals of the \ReinitPP in pseudocode. When an MPI process executes
\verb|MPI_Reinit|, it installs a \emph{signal handler} for the signal
\verb|SIGREINIT|, which aliases \verb|SIGUSR1| in our implementation. Also,
\verb|MPI_Reinit| sets a non-local goto point using the POSIX function
\verb|setjmp()|. The signal handler of
\verb|SIGREINIT| simply calls \verb|longjmp()| to return execution of survivor
processes to this goto point.  Rolled back survivor processes discard any
previous MPI state and block on a ORTE-level barrier. This barrier replicates
the implicit barrier present in \verb|MPI_Init| to synchronize with re-spawned
processes joining the computation. After the barrier, survivor processes
re-initialize the world communicator and call the function \verb|foo| to resume
computation. Re-spawned processes initialize the world communicator as part of
the MPI initialization procedure of \verb|MPI_Init| and go through 
\verb|MPI_Reinit| to install the signal handler, set the goto point, and
lastly call the user-defined function to resume computation.


\subsubsection{Application Recovery}

Application recovery includes the actions needed at the
application-level to resume computation.
Any additional MPI state besides the repaired world communicator, such
as sub-communicators, must be re-created by the application's MPI
processes. Also, it is expected that each process loads the latest consistent
checkpoint to continue computing. Checkpointing lays within the
responsibility of the application developer. In the next section, we
discuss the scope and implications of our implementation.

\subsubsection{Discussion}
In this implementation, the scope of fault tolerance is to support
recovery from failures \emph{happening after} \verb|MPI_Reinit| has been
called by all MPI processes. This is because \verb|MPI_Reinit| must
install signal handlers and set the roll-back point on all MPI
processes. This is sufficient for a large coverage of failures since
execution time is dominated by the main computational loop. In the
case a failure happens before the call to
\verb|MPI_Reinit|, the application falls back to the default action of
aborting execution. Nevertheless, the design of \ReinitPP is not limited by
this implementation choice. A possible approach instead of aborting,
which we leave as future work, is to treat any MPI processes that
have not called \verb|MPI_Reinit| as if failed and re-execute them.

Furthermore, signaling \verb|SIGREINIT| for rolling back survivor MPI
processes asynchronously interrupts execution.  In our
implementation, we render the MPI runtime library \emph{signal and
roll-back safe} by using masking to defer signal handling until a safe
point, i.e., avoid interruption when locks are held or data structures
are updating. Since application code is out of our control, \ReinitPP
requires the application developer to program the
application as signal and roll-back safe. A possible enhancement is to
provide an interface for installing cleanup handlers, proposed in
earlier designs of Reinit~\cite{Laguna:2014:EUF:2642769.2642775}, so
that application and library developers can install routines to  reset
application-level state on recovery. Another approach is to make
recovery synchronous, by extending the \ReinitPP interface to include a
function that tests whether a fault has been detected and trigger roll
back. The developer may call this function at safe points during
execution for recovery. We leave both those enhancements as future work,
noting that the existing interface is sufficient for performing our
evaluation.

\section{Experimentation Setup}
\label{sec:setup}

This section provides detailed information on the experimentation setup,
the recovery approaches used for comparisons, the proxy applications and
their configurations, and the measurement methodology.

\subsubsection{Recovery approaches}
Experimentation includes the following recovery approaches:
\begin{itemize}
  \item \emph{CR}, which implements the typical approach of
    immediately restarting an application after execution aborts due to a failure. 
  \item \emph{ULFM}, by using its latest revision
    based on the Open MPI runtime v4.0.1 (4.0.1ulfm2.1rc1). 
  \item \emph{\ReinitPP}, which is our own implementation of Reinit, based on OpenMPI
    runtime v4.0.0.
\end{itemize}

\subsubsection{Emulating failures}

Failures are emulated through fault injection.  We opt for random fault
injection to emulate the occurrence of random faults, e.g., soft errors
or failures of hardware components, that lead to a crash failure.
Specifically, for process failures, we instrument applications so that
at a random iteration of the main computational loop, a random MPI
process suicides by raising the signal \verb|SIGKILL|. The random
selection of iteration and MPI process is the same for every recovery
approach.  For node failures, the method is similar, but instead of
itself, the MPI process sends the signal \verb|SIGKILL| to its parent
daemon, thus kills the daemon and by extension all its children
processes. In experimentation, we inject a \emph{single} MPI process
failure or a \emph{single} node failure.

\subsubsection{Applications}
\begin{table}[t]
  \setlength{\tabcolsep}{8pt}
  \centering
  \caption{Proxy applications and their configuration}
  \begin{tabular}{lccc}
  \toprule
  \textbf{Application} & \textbf{Input} & \textbf{No. ranks}\\
  \midrule
  CoMD & -i4 -j2 -k2 & 16, 32, 64, 128, 256, 512, 1024 \\
      & -x 80 -y 40 -z 40 -N 20 & & \\
  HPCCG & 64 64 64 & 16, 32, 64, 128, 256, 512, 1024 \\
  LULESH & -i 20 -s 48 & 8, 64, 512 \\
  \bottomrule
\end{tabular}

  \label{tab:apps}
\end{table}
We experiment with three benchmark applications that represent different
HPC domains: \emph{CoMD} for molecular dynamics, \emph{HPCCG} for
iterative solvers, and \emph{LULESH} for multi-physics computation. The
motivation is to investigate global-restart recovery on a wide range of
applications and evaluate any performance differences.
Table~\ref{tab:apps} shows information on the proxy applications and
scaling of their deployed number of ranks.  Note \emph{LULESH} requires
a cube number of ranks, thus the trimmed down experimentation space. The
deployment configuration has 16 ranks per node, so the smallest
deployment comprises of one node while the largest one spans 64 nodes
(1024 ranks).  Application execute in \emph{weak scaling} mode -- for
\emph{CoMD} we show its input only 16 ranks and change it accordingly.
We extend applications to implement global-restart with \ReinitPP or ULFM,
to store a checkpoint after every iteration of their main computational
loop and load the latest checkpoint upon recovery. 

\subsubsection{Checkpointing}

\begin{table}[t]
  \centering
  \setlength{\tabcolsep}{12pt}
  \caption{Checkpointing per recovery and failure}
  \begin{tabular}{cccc}
  \toprule
  \textbf{Failure} & \multicolumn{3}{c}{\textbf{Recovery}} \\
  \midrule
          & CR & ULFM & Reinit \\
  \cmidrule{2-4}
  \emph{process} & file & memory & memory \\
  \midrule
  \emph{node}    & file & file & file \\

  \bottomrule
\end{tabular}

  \label{tab:checkpoints}
\end{table}

For evaluation purposes, we implement our own, simple checkpointing
library that supports saving and loading application data using
in-memory and file checkpoints.
Table~\ref{tab:checkpoints} summarizes checkpointing per recovery
approach and failure type. In detail, we implement two types of
checkpointing: \emph{file} and \emph{memory}. For file checkpointing,
each MPI process stores a checkpoint to globally accessible permanent
storage, which is the networked, parallel filesystem Lustre available in
our cluster. For memory checkpointing, an MPI process stores a
checkpoint both locally in its own memory and remotely to the memory of
a \emph{buddy}~\cite{Zheng:2006:PEA:1131322.1131340,6264677} MPI
process, which in our implementation is the (cyclically) next MPI
process by rank. This memory checkpointing implementation is applicable
only to single process failures since multiple process failures or a
node failure can wipe out both local and buddy checkpoints for the
failed MPI processes. CR necessarily uses file checkpointing since
re-deploying the application requires permanent storage to retrieve
checkpoints.

\subsubsection{Statistical evaluation}
For each proxy application and configuration we perform 10 independent
measurements.  Each measurement counts the total execution time of the
application breaking it down to time needed for writing checkpoints,
time spent during MPI recovery, time reading a checkpoint after a
failure, and the pure application time executing the computation.
Any confidence intervals shown correspond to a 95\% confidence
level and are calculated based on the t-distribution to avoid
assumptions on the sampled population's distribution.

\section{Evaluation}
\label{sec:evaluation}

For the evaluation we compare CR, \ReinitPP and ULFM for both process and
node failures. Results provide insight on the performance of each of
those recovery approaches implementing global-restart and reveal the
reasons for their performance differences.

\begin{figure}[t]
  \centering
  \begin{subfigure}[t]{.45\textwidth}
    \includegraphics[width=1.0\textwidth]{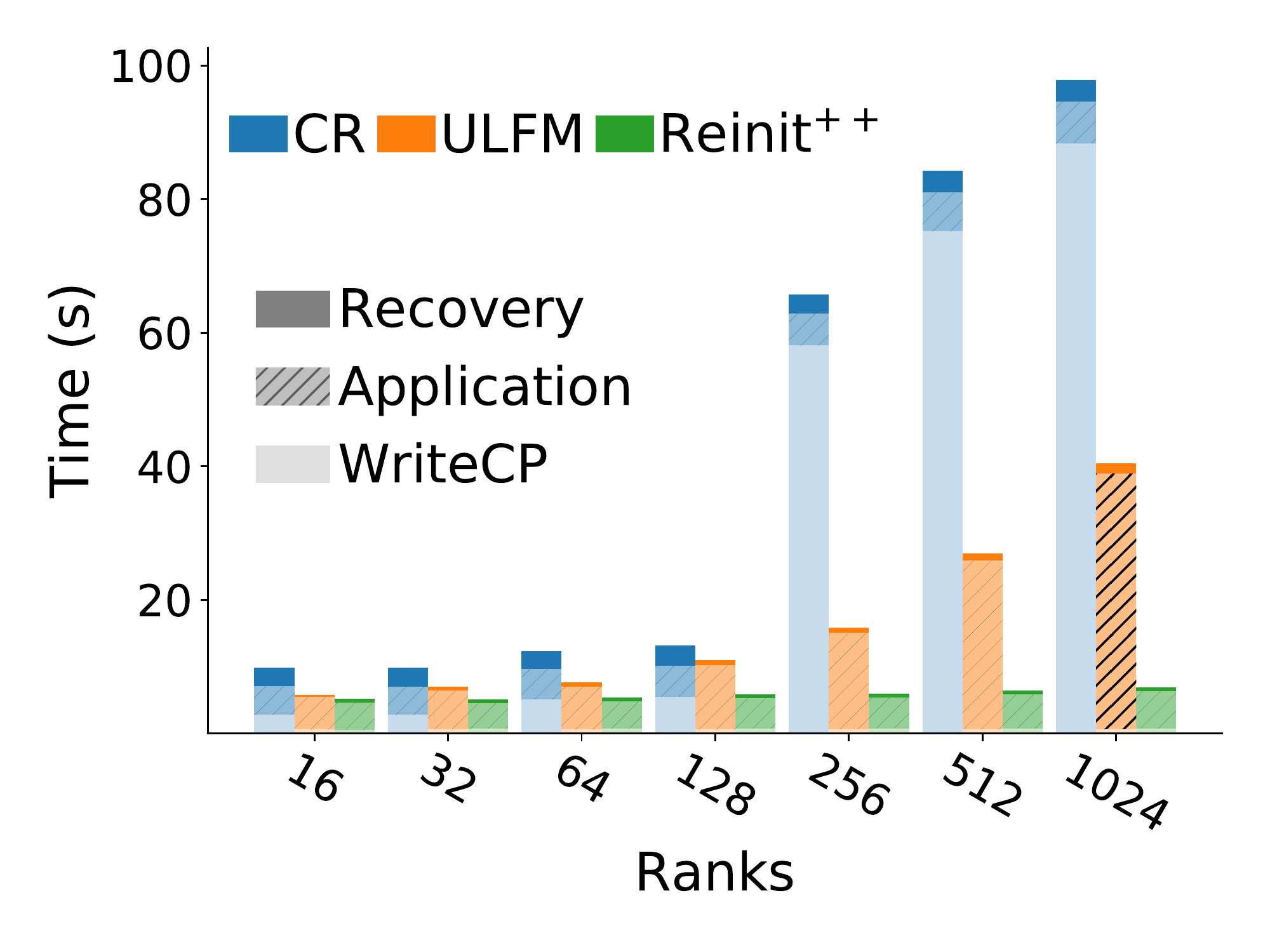}
    \caption{CoMD}
  \end{subfigure}
  \begin{subfigure}[t]{.45\textwidth}
    \includegraphics[width=1.0\textwidth]{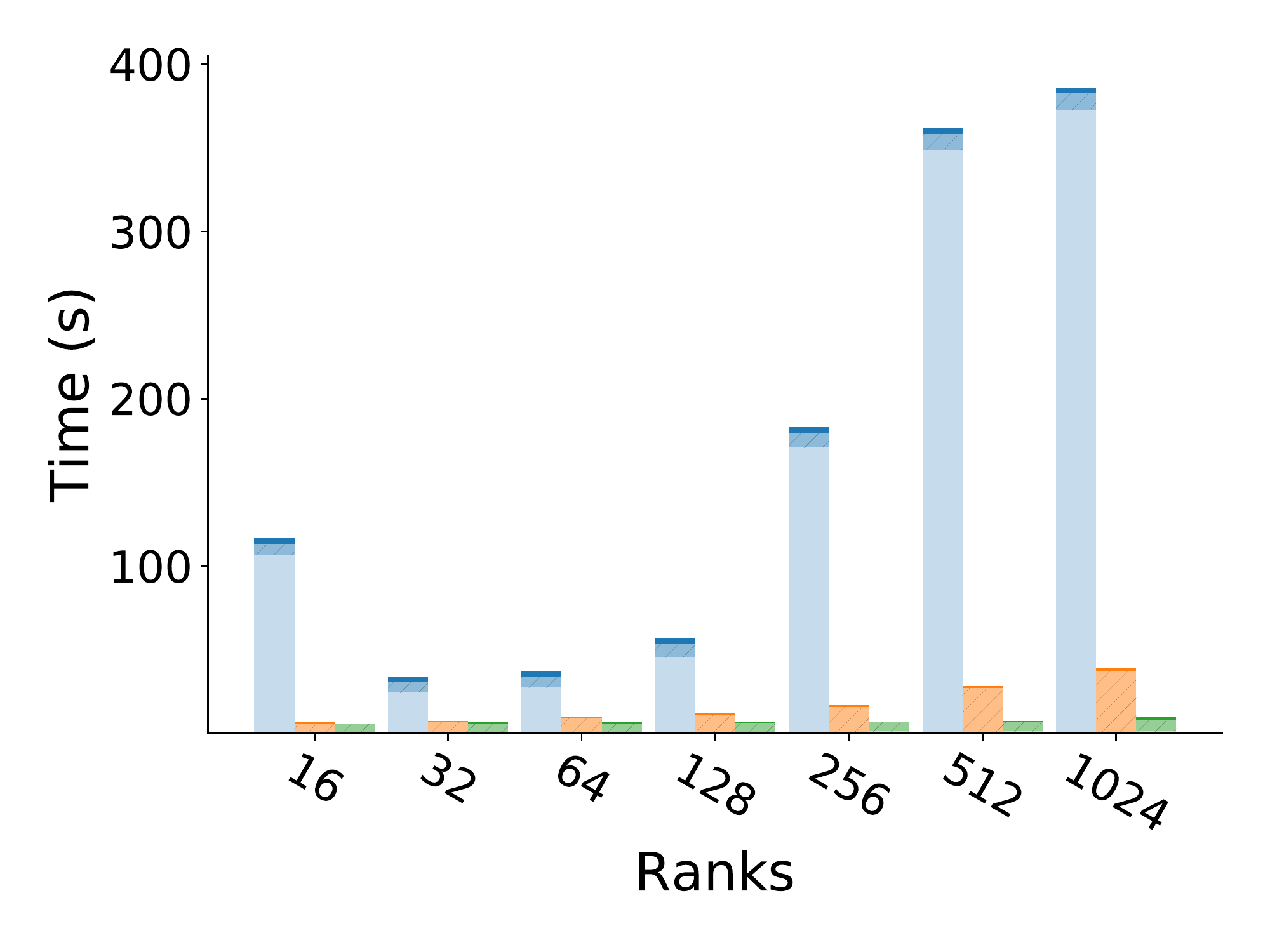}
    \caption{HPCCG}
    \label{fig:tot:procfi:hpccg}
  \end{subfigure}
  \begin{subfigure}[t]{.45\textwidth}
    \includegraphics[width=1.0\textwidth]{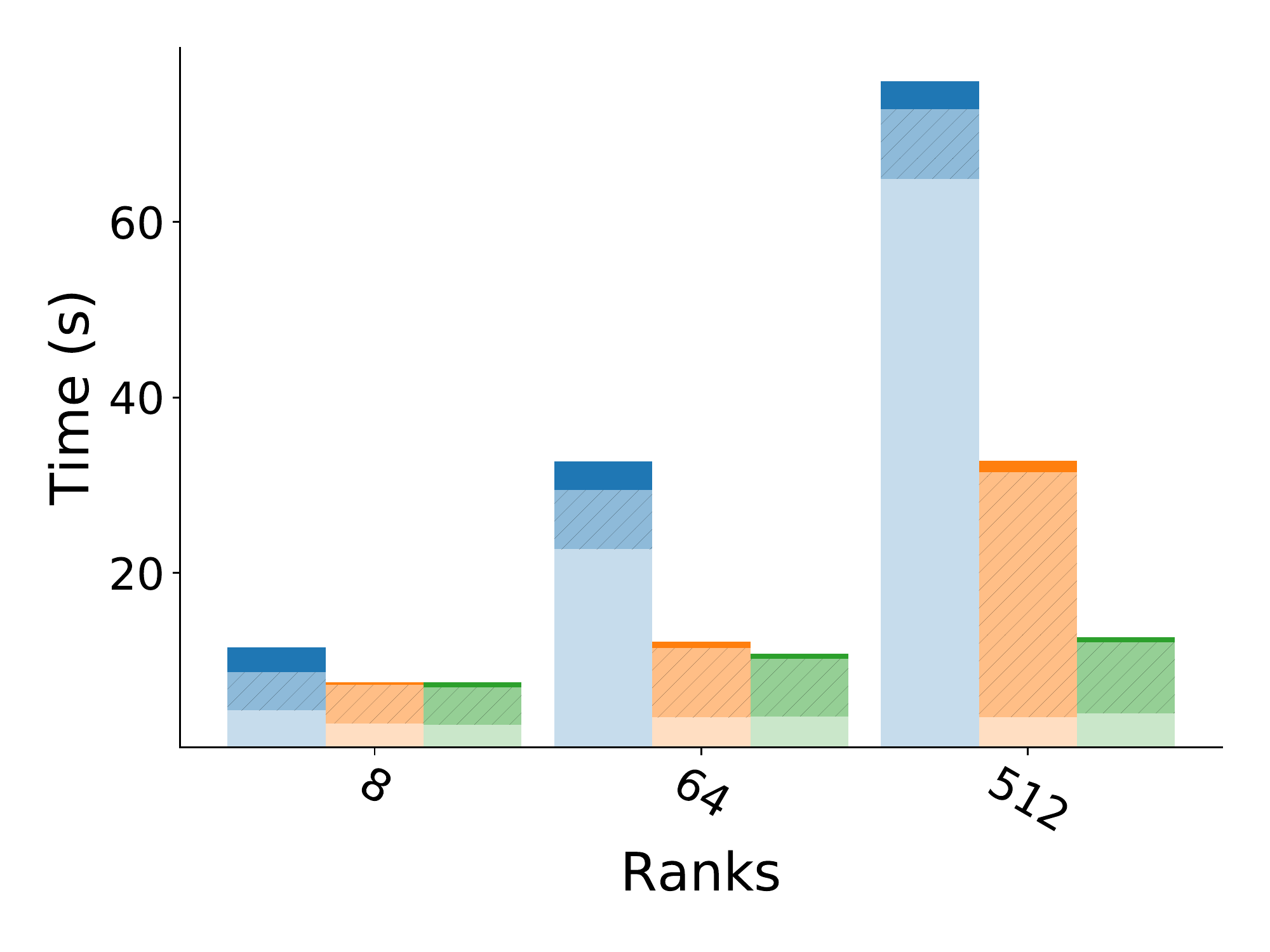}
    \caption{LULESH}
  \end{subfigure}
  \caption{Total execution time breakdown recovering from a process failure}
  \label{fig:tot:procfi}
\end{figure}

\subsection{Comparing total execution time on a process failure}

Figure~\ref{fig:tot:procfi} shows average total execution time for
process failures using file checkpointing for CR and memory
checkpointing for \ReinitPP and ULFM. The plot breaks down time to
components of writing checkpoints, MPI recovery, and pure application
time. Reading checkpoints occurs one-off after a failure and has
negligible impact, in the order of tens of milliseconds, thus it is
omitted. 

The first observation is that \ReinitPP scales excellently compared to both
CR and ULFM, across all programs. CR has the worse performance,
increasingly so with more ranks.  The reason is the limited scaling of
writing checkpoints to the networked filesystem. By contrast, ULFM and
\ReinitPP use memory checkpointing, spending minimal time writing
checkpoints.  Interestingly, ULFM scales worse than \ReinitPP; we believe
that the reason is that it inflates pure application execution time,
which we illustrate in the next section.  Further, in the following
sections, we remove checkpointing overhead from the analysis to
highlight the performance differences of the different recovering
approaches.


\begin{figure}[t]
  \centering
  \begin{subfigure}[t]{.328\textwidth}
    \includegraphics[width=1.0\textwidth]{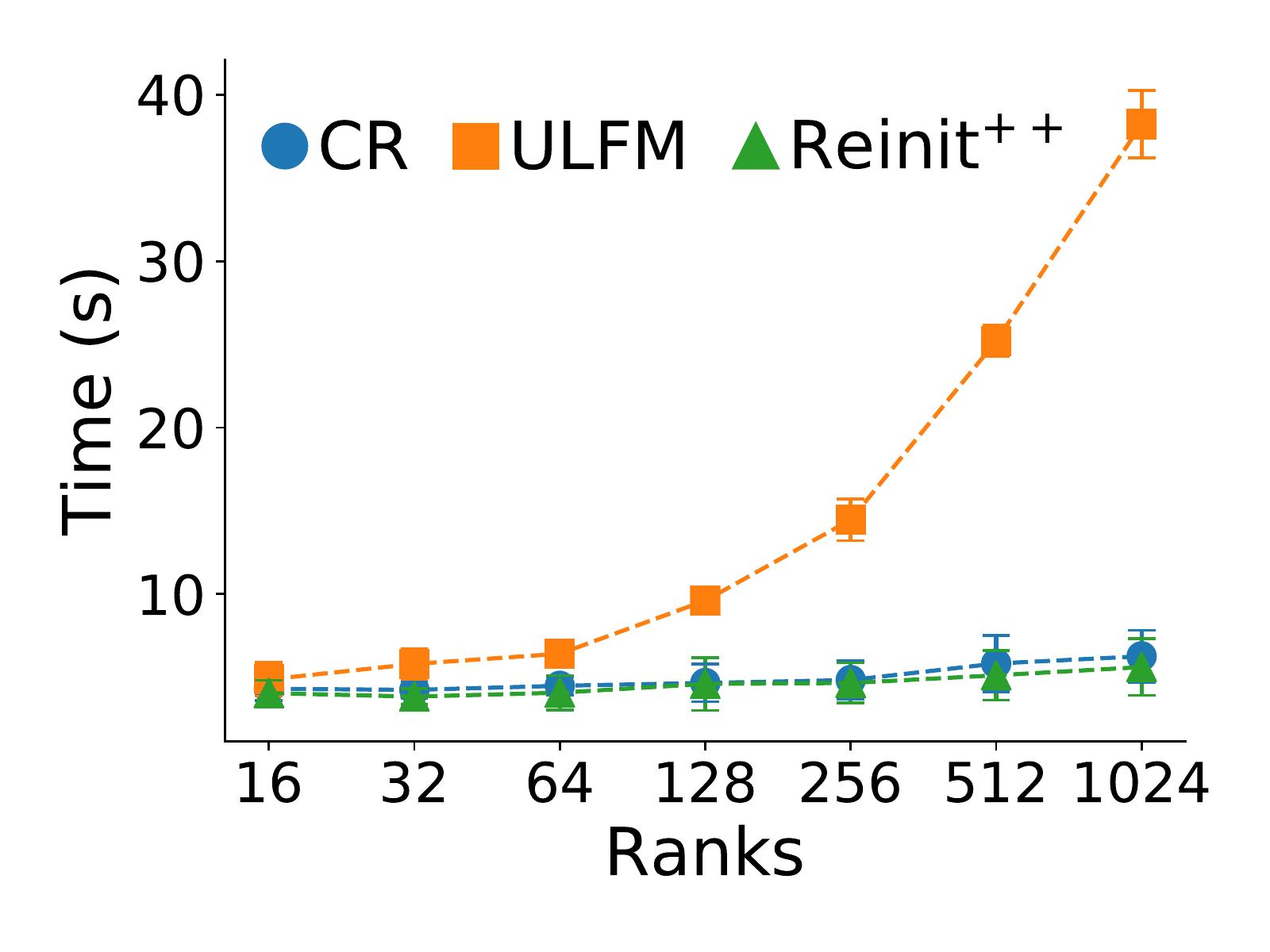}
    \caption{CoMD}
  \end{subfigure}
  \begin{subfigure}[t]{.328\textwidth}
    \includegraphics[width=1.0\textwidth]{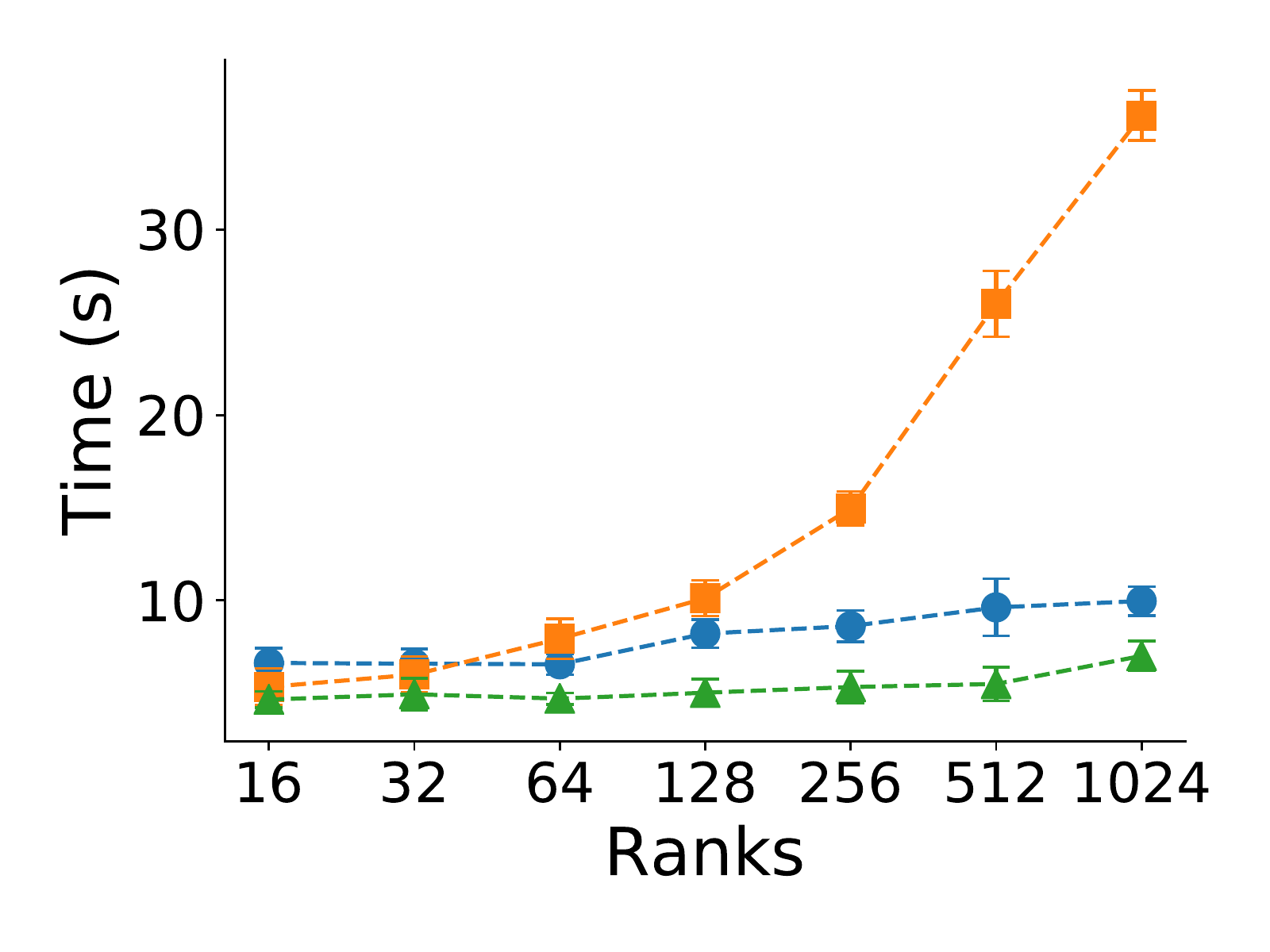}
    \caption{HPCCG}
  \end{subfigure}
  \begin{subfigure}[t]{.328\textwidth}
    \includegraphics[width=1.0\textwidth]{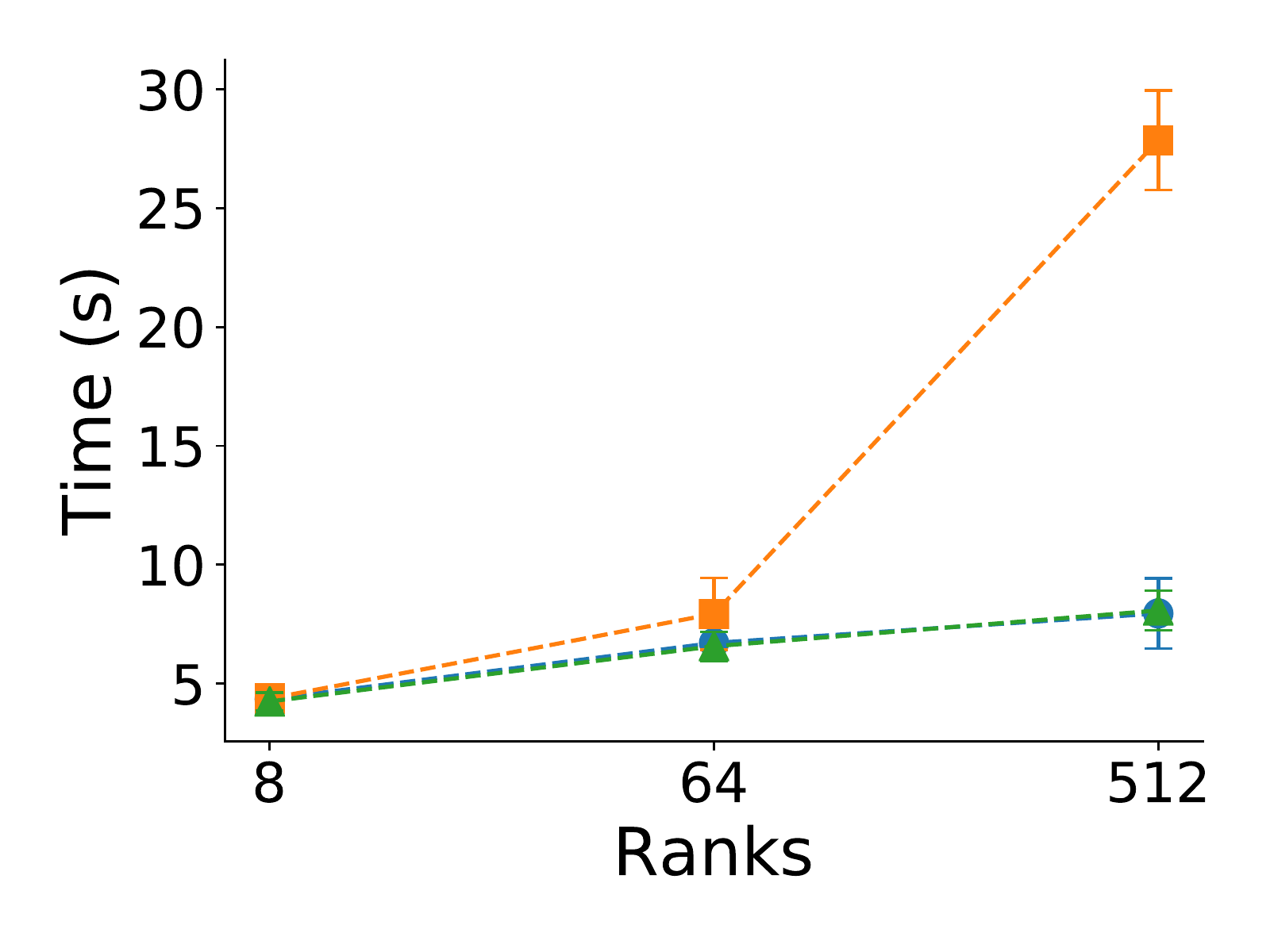}
    \caption{LULESH}
  \end{subfigure}
  \caption{Scaling of pure application time}
  \label{fig:app:procfi}
\end{figure}

\subsection{Comparing pure application time under different recovery approaches}

Figure~\ref{fig:app:procfi} shows the pure application time, without
including reading/writing checkpoints or MPI recovery. We observe that
application time is on par for CR and \ReinitPP, and that all
applications scale weakly well on up to 1024 ranks. CR and \ReinitPP
do not interfere with execution, thus they have no impact on application
time, which is on par to the fault-free execution time of the proxy
applications. However, in ULFM, application time grows
significantly as the number of ranks increases.
ULFM extends MPI with an always-on,
periodic heartbeat mechanism~\cite{doi:10.1177/1094342017711505} to detect failures and also modifies
communication primitives for fault tolerant operation. Following from
our measurements, those extensions noticeably increase the original application
execution time. However, it is inconclusive whether this is a result of
the tested prototype implementation or a systemic trade-off. Next, we
compare the MPI recovery times among all the approaches.


\begin{figure}[t]
  \centering
  \begin{subfigure}[t]{.328\textwidth}
    \includegraphics[width=1.0\textwidth]{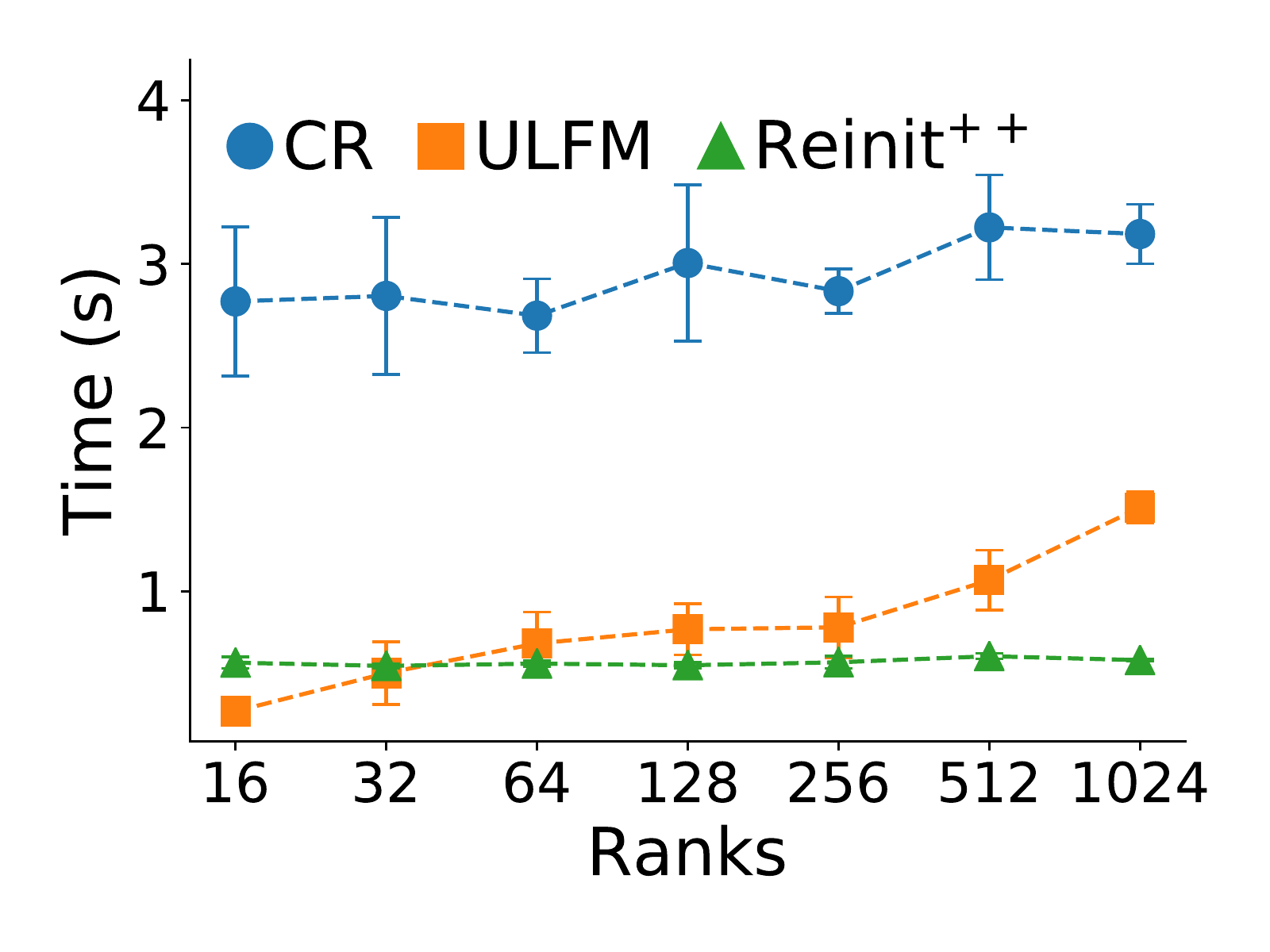}
    \caption{CoMD}
  \end{subfigure}
  \begin{subfigure}[t]{.328\textwidth}
    \includegraphics[width=1.0\textwidth]{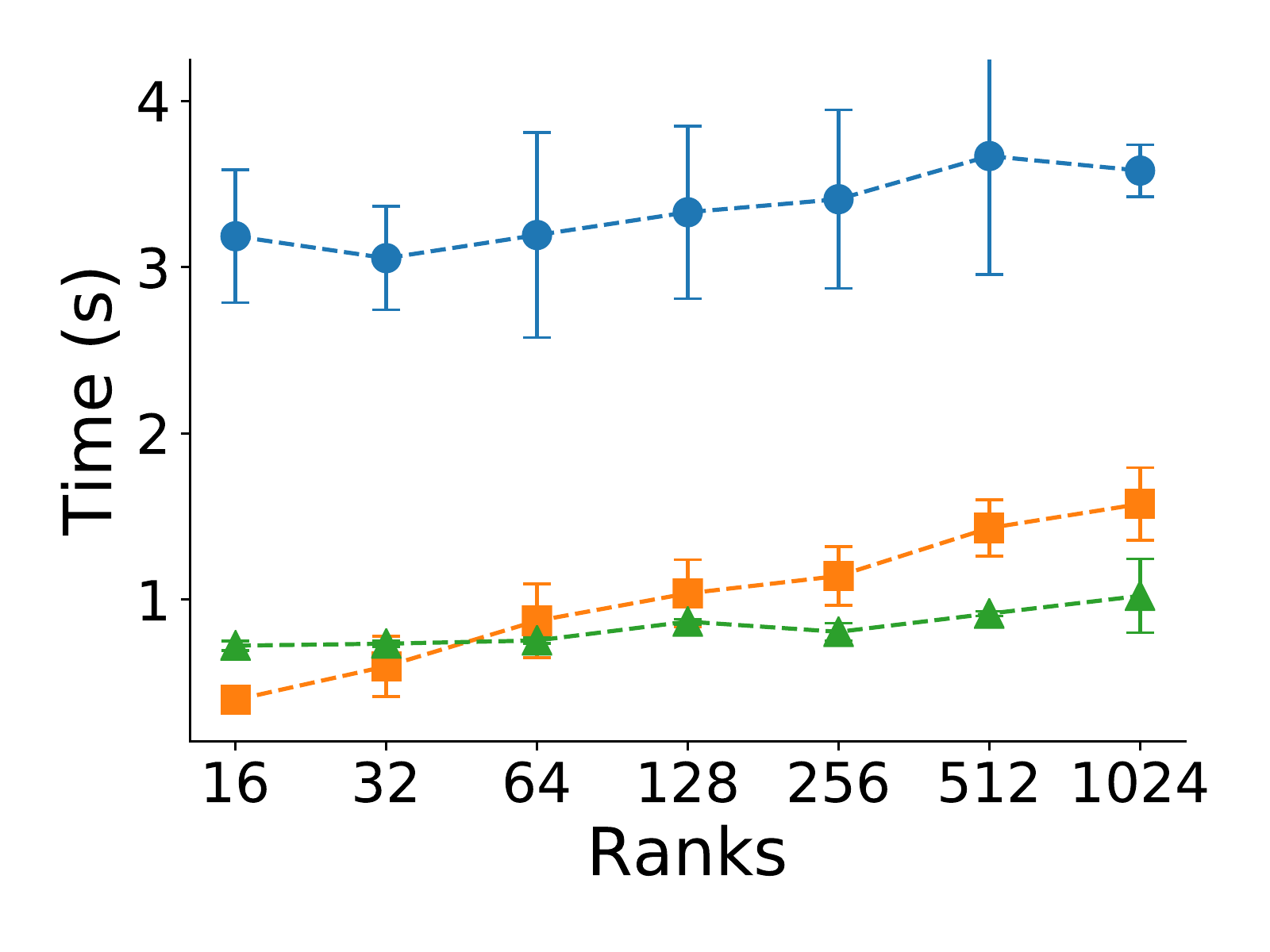}
    \caption{HPCCG}
  \end{subfigure}
  \begin{subfigure}[t]{.328\textwidth}
    \includegraphics[width=1.0\textwidth]{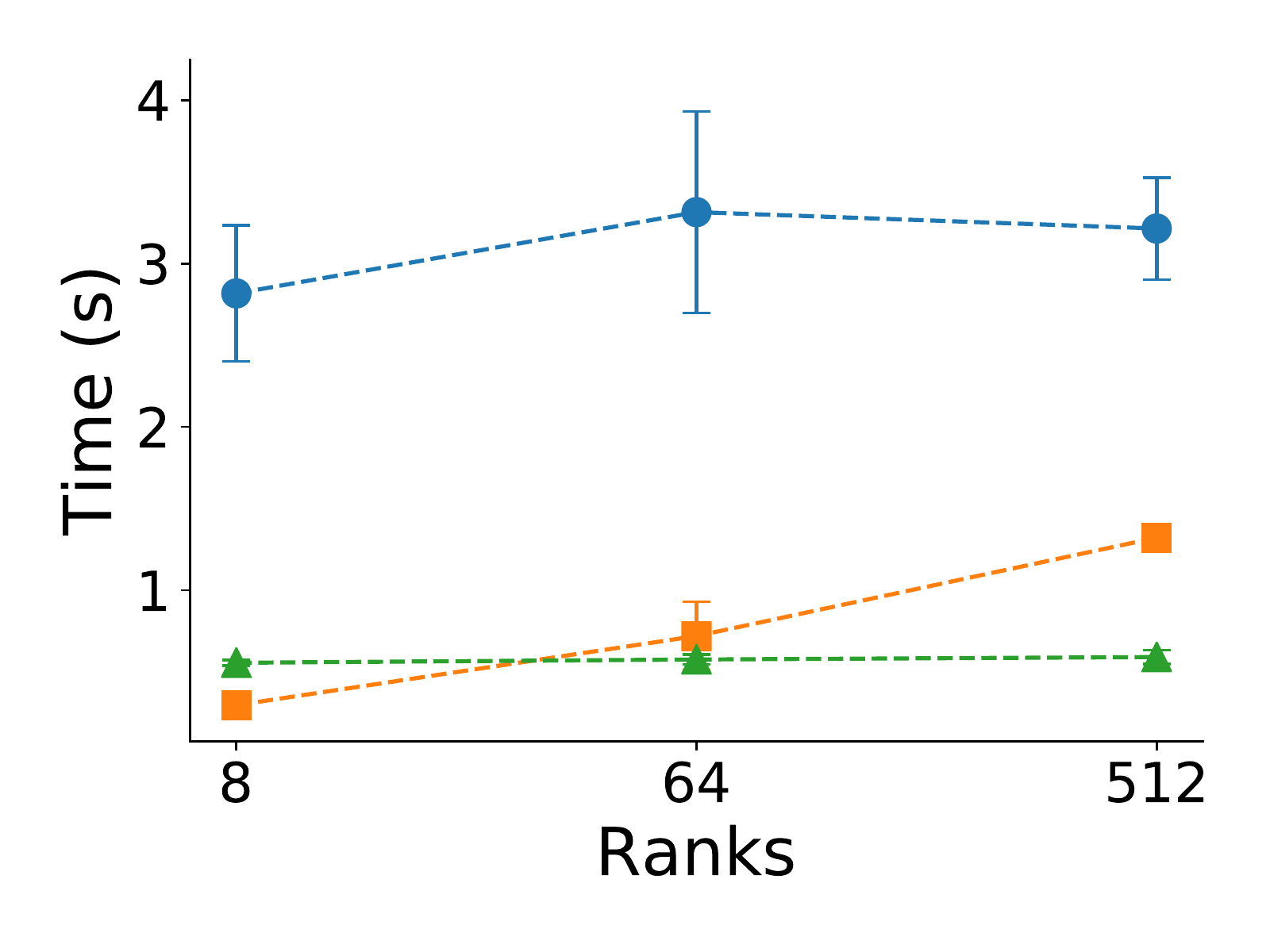}
    \caption{LULESH}
  \end{subfigure}
  \caption{Scaling of MPI recovery time recovering from a process failure}
  \label{fig:recov:procfi}
\end{figure}

\subsection{Comparing MPI recovery time recovering from  a process failure}

Though checkpointing saves application’s computation time, reducing MPI
recovery time saves overhead from restarting. This overhead is
increasingly important the larger the deployment and the higher the
fault rate. In particular, figure~\ref{fig:recov:procfi} shows the
scaling of time required for MPI recovery across all programs and
recovery approaches, again removing any overhead for checkpointing to
focus on the MPI recovery time. As expected, MPI recovery time depends
only on the number of ranks, thus times are similar among different
programs for the same recovery approach.  Commenting on scaling, CR and
\ReinitPP scale excellently, requiring almost constant time for MPI
recovery regardless the number of ranks.  However, CR is about 6$\times$
slower, requiring around 3 seconds to tear down execution and re-deploy
the application, whereas \ReinitPP requires about 0.5 second to propagate
the fault, re-initialize survivor processes and re-spawn the failed
process. ULFM has on par recovery time with \ReinitPP up to 64 ranks, but
then its time increases being up to 3$\times$ slower than \ReinitPP for
1024 ranks.  ULFM requires multiple collective operations among all MPI
processes to implement global-restart (shrink the faulty communicator,
spawn a new process, merge it to a new communicator).  By contrast,
\ReinitPP implements recovery at the MPI runtime layer requiring fewer
operations and confining collective communication only between root and
daemon processes.


\begin{figure}[t]
  \centering
  \begin{subfigure}[t]{.328\textwidth}
    \includegraphics[width=1.0\textwidth]{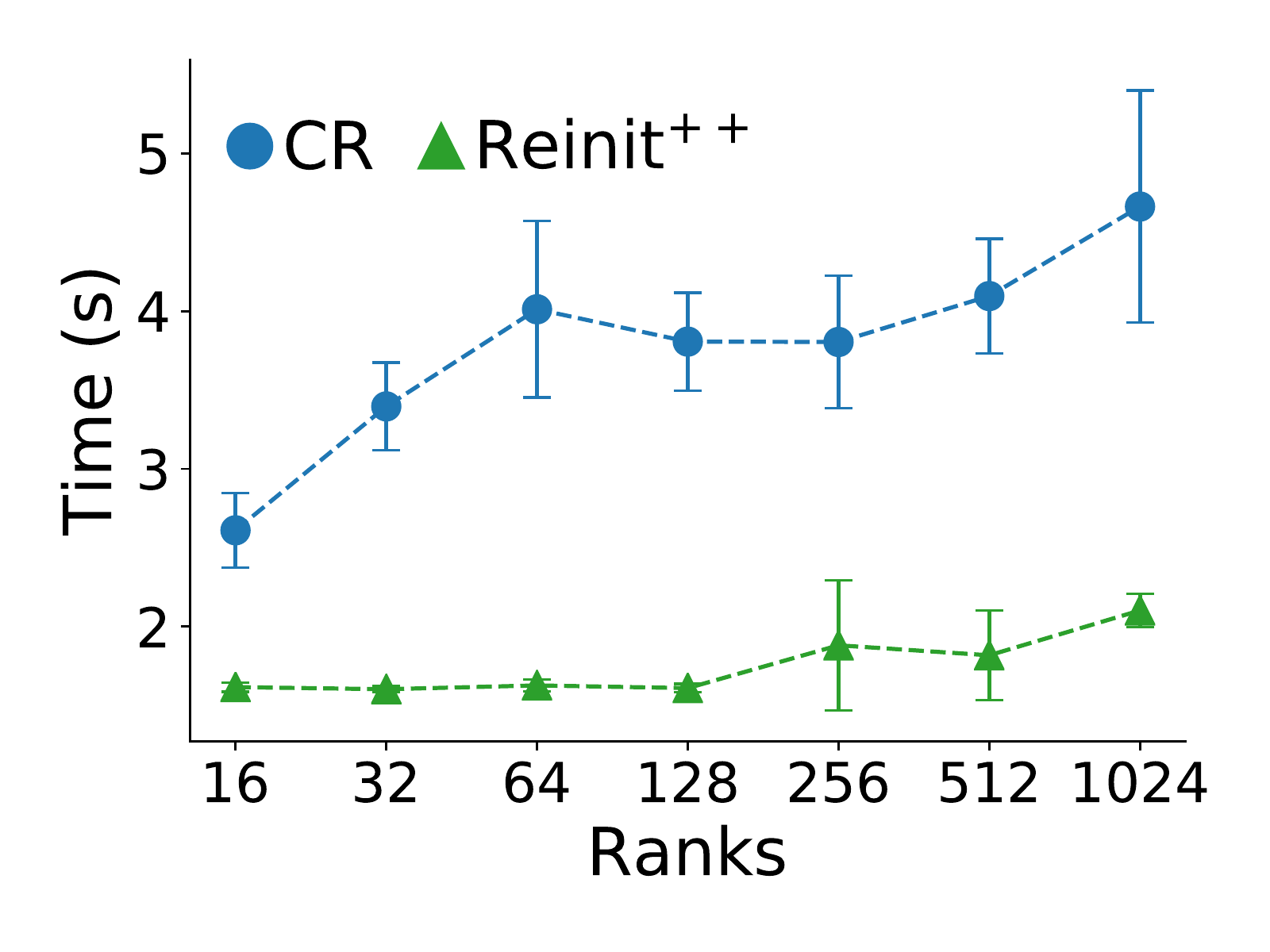}
    \caption{CoMD}
  \end{subfigure}
  \begin{subfigure}[t]{.328\textwidth}
    \includegraphics[width=1.0\textwidth]{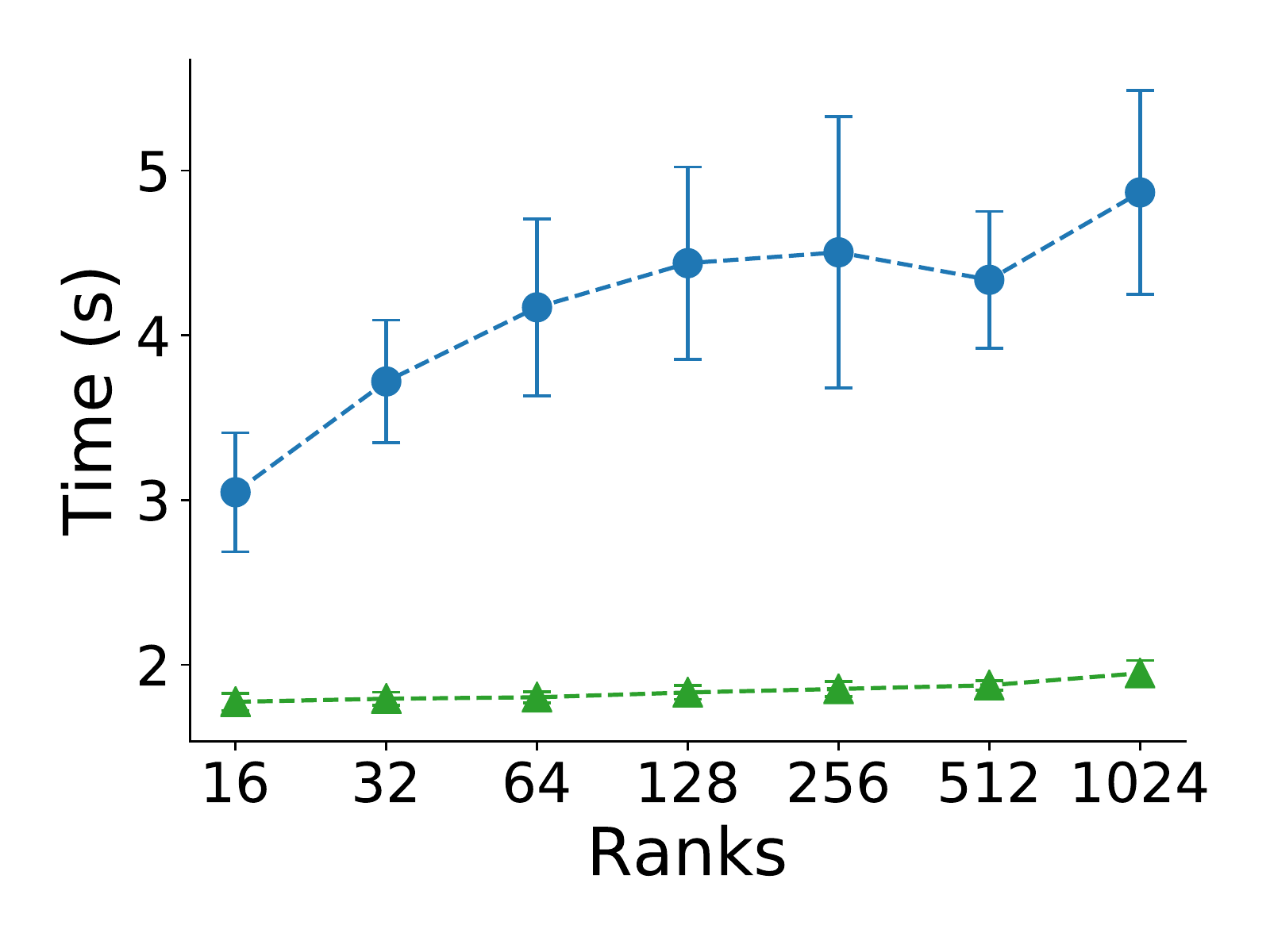}
    \caption{HPCCG}
  \end{subfigure}
  \begin{subfigure}[t]{.328\textwidth}
    \includegraphics[width=1.0\textwidth]{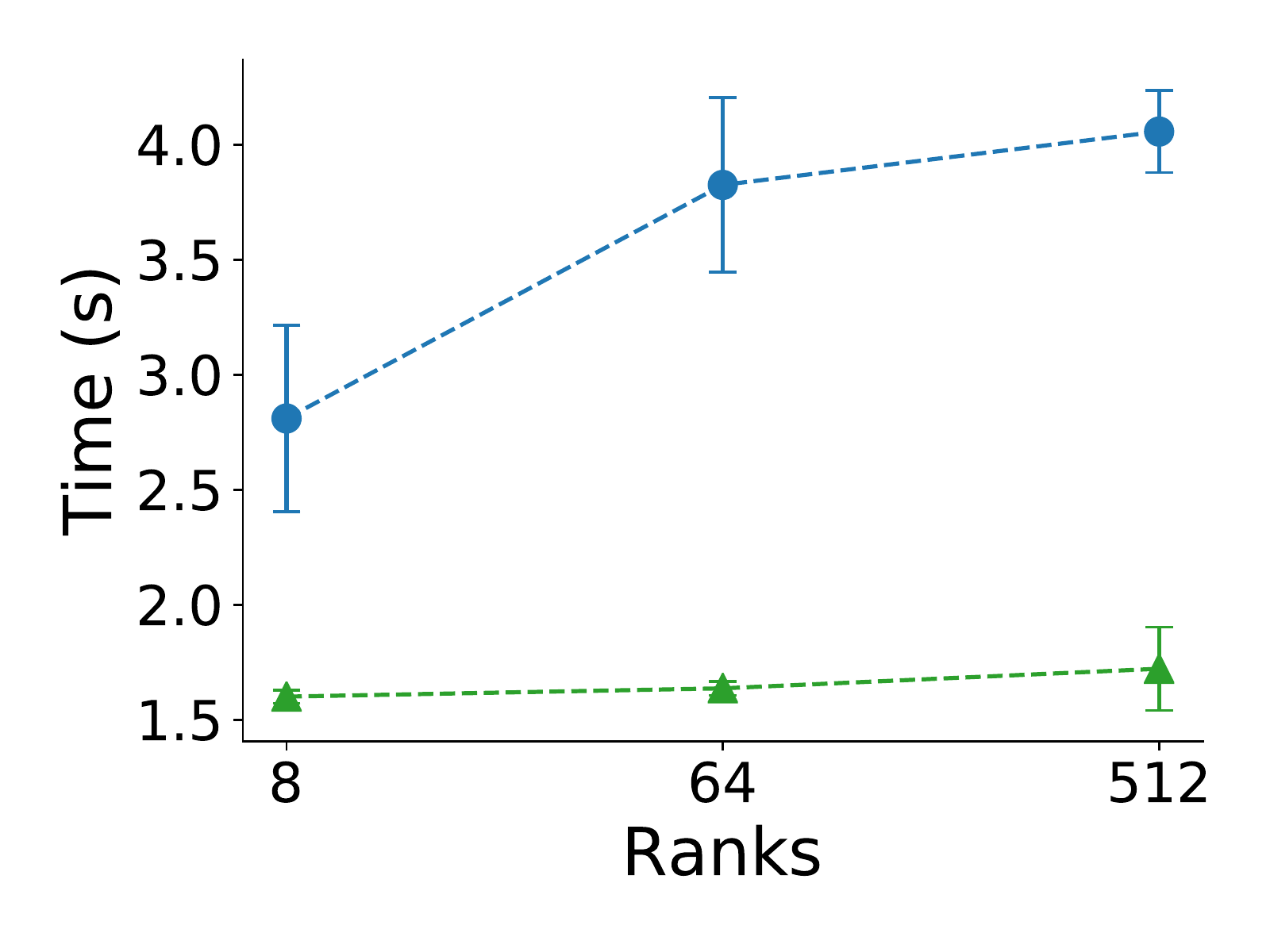}
    \caption{LULESH}
  \end{subfigure}
  \caption{Scaling of MPI recovery time recovering from a node failure}
  \label{fig:recov:nodefi}
\end{figure}

\subsection{Comparing MPI recovery time recovering from a node failure}

This comparison for a node failure includes only CR and \ReinitPP, since
the prototype implementation of ULFM faced robustness issues (hanging or
crashing) and did not produce measurements. Also, since both CR and
\ReinitPP use file checkpointing and do not interfere with pure application
time, we present only results for MPI recovery times, shown in
figure~\ref{fig:recov:nodefi}. Both CR and \ReinitPP scale very well with
almost constant times, as they do for a process failure. However, in
absolute values, \ReinitPP has a higher recovery time of about 1.5 seconds
for a node failure compared to 0.5 seconds for a process failure. This
is because recovering from a node failure requires extra work to select
the least loaded node and spawn all the MPI processes of the failed
node.  Nevertheless, recovery with \ReinitPP is still about 2$\times$
faster than with CR.

\section{Related Work}
\label{sec:related}

Checkpoint-Restart~\cite{hargrove2006berkeley,sankaran2005lam,adam2019checkpoint,subasi2018unified,wang2018fault,cao2016system,adam2018transparent,kohl2019scalable}
is the most common approach to recover an MPI application after a
failure. CR requires substantial development effort to identify which
data to checkpoint and may have significant overhead. Thus, many efforts
attempt to make checkpointing easier to adopt and render it fast and
storage efficient.  We briefly discuss them here.

Hargrove and Duell~\cite{hargrove2006berkeley} implement the
system-level CR library Berkeley Lab Checkpoint/Restart (BLCR) library
to automatically checkpoint applications by extending the Linux kernel.
Bosilca et al.~\cite{bosilca2002mpich} integrate an uncoordinated,
distributed checkpoint/roll-back system in the MPICH runtime to
automatically support fault tolerance for node failures.  Furthermore.
Sankaran et al.~\cite{sankaran2005lam} integrate Berkeley Lab BLCR
kernel-level C/R to the LAM implementation of MPI.  Adam et
al.~\cite{adam2019checkpoint}, SCR~\cite{6494566}, and
FTI~\cite{6114441} propose asynchronous, multi-level checkpointing
techniques that significantly improve checkpointing performance.
Shahzad et al.~\cite{shahzad2018craft} provide an extensive interface
that simplifies the implementation of application-level
checkpointing and recovery. Advances in checkpointing are beneficial not
only for CR but for other MPI fault tolerance approaches, such as ULFM
and Reinit. Though making checkpointing faster resolves this bottleneck,
the overhead of re-deploying the full application remains.

ULFM~\cite{bland2013post,bland2015lessons} is the state-of-the-art MPI
fault tolerance approach, pursued by the MPI Fault Tolerance
Working Group.  ULFM extends MPI with interfaces to shrink or revoke
communicators, and fault-tolerant collective consensus. The application
developer is responsible for implementing recovery using those
operations, choosing the type of recovery best suited for its
application.  A collection of works on
ULFM~\cite{losada2017resilient,pauli2014fault,hori2015sliding,katti2015scalable,herault2015practical,bouteiller2015plan,Laguna:2014:EUF:2642769.2642775}
has investigated the applicability of ULFM and benchmarked
individual operations of it.  Bosilca et
al.~\cite{bosilca2016failure,doi:10.1177/1094342017711505} and Katti et
al.~\cite{katti2018epidemic} propose efficient fault detection
algorithms to integrate with ULFM.  
Teranishi et
al.~\cite{teranishi2014toward} use spare processes to replace
failed processes for local recovery so as to accelerate recovery
of ULFM.  
Even though ULFM gives
flexibility to developers to implement any type of recover, 
it requires significant developer effort to refactor the application.
Also, implementing ULFM has been
identified by previous work~\cite{teranishi2014toward,gamell2015local}
to suffer from scalability issues, as our experimentation shows too.
Fenix~\cite{Gamell2014Fenix} provides a simplified
abstraction layer atop ULFM 
to implement global-restart recovery. However, we choose to
directly use ULFM since it already provides a straightforward, prescribed solution for
implementing global-restart.

Reinit~\cite{doi:10.1177/1094342015623623,doi:10.1002/cpe.4863} 
is an alternative solution that supports only global-restart recovery
and provide an easy to use interface to developers. Previous designs and
implementations of Reinit have limited applicability because they
require modifying the job scheduler and its interface with the MPI
runtime. We present \ReinitPP, a new design and implementation of Reinit
using the latest Open MPI runtime and thoroughly evaluate it.

Lastly, Sultana et al.~\cite{SULTANA20191} propose MPI stages to reduce
the overhead of global-restart recovery by checkpointing MPI state, so
that rolling back does not have to re-create it. While this approach is
interesting, it is still in proof-of-concept status. How to maintain 
consistent checkpoints of MPI state across all MPI processes,
and doing so fast and efficiently, is still an open-problem.

\section{Conclusion}
\label{sec:conclusion}

We have presented \ReinitPP, a new design and implementation of the global-restart
approach of Reinit. \ReinitPP recovers from both process and node crash
failures, by spawning new processes and mending the world communicator,
requiring from the programmer only to provide a rollback point in
execution and have checkpointing in place.  Our extensive evaluation 
comparing with the state-of-the-art approaches Checkpoint-Restart
(CR) and ULFM shows that \ReinitPP scales excellently as the number of ranks
grows, achieving almost constant recovery time, being up to 6$\times$
faster than CR and up to 3$\times$ faster than ULFM.  For future work,
we plan to expand Reinit for supporting more recovery strategies besides
global-restart, including shrinking recovery and forward recovery
strategies, to maintain its implementation, and expand the
experimentation with more applications and larger deployments.

\section*{Acknowledgments}
The authors would like to thank the anonymous referees for their
valuable comments and helpful suggestions.  This work was performed
under the auspices of the U.S.~Department of Energy by Lawrence
Livermore National Laboratory under contract DEAC52-07NA27344
(LLNL-CONF-800061).

\bibliographystyle{splncs04}
{\tiny \bibliography{references}}

\begin{thebibliography}{10}
\providecommand{\url}[1]{\texttt{#1}}
\providecommand{\urlprefix}{URL }
\providecommand{\doi}[1]{https://doi.org/#1}

\bibitem{adam2018transparent}
Adam, J., Besnard, J.B., Malony, A.D., Shende, S., P{\'e}rache, M., Carribault,
  P., Jaeger, J.: Transparent high-speed network checkpoint/restart in mpi. In:
  Proceedings of the 25th European MPI Users' Group Meeting. p.~12 (2018)

\bibitem{adam2019checkpoint}
Adam, J., Kermarquer, M., Besnard, J.B., Bautista-Gomez, L., P{\'e}rache, M.,
  Carribault, P., Jaeger, J., Malony, A.D., Shende, S.: Checkpoint/restart
  approaches for a thread-based mpi runtime. Parallel Computing  \textbf{85},
  204--219 (2019)

\bibitem{6114441}
{Bautista-Gomez}, L., {Tsuboi}, S., {Komatitsch}, D., {Cappello}, F.,
  {Maruyama}, N., {Matsuoka}, S.: Fti: High performance fault tolerance
  interface for hybrid systems. In: SC '11: Proceedings of 2011 International
  Conference for High Performance Computing, Networking, Storage and Analysis.
  pp. 1--12 (Nov 2011). \doi{10.1145/2063384.2063427}

\bibitem{bland2013post}
Bland, W., Bouteiller, A., Herault, T., Bosilca, G., Dongarra, J.: Post-failure
  recovery of mpi communication capability: Design and rationale. The
  International Journal of High Performance Computing Applications
  \textbf{27}(3),  244--254 (2013)

\bibitem{bland2015lessons}
Bland, W., Lu, H., Seo, S., Balaji, P.: Lessons learned implementing user-level
  failure mitigation in mpich. In: 2015 15th IEEE/ACM International Symposium
  on Cluster, Cloud and Grid Computing (2015)

\bibitem{bosilca2002mpich}
Bosilca, G., Bouteiller, A., Cappello, F., Djilali, S., Fedak, G., Germain, C.,
  Herault, T., Lemarinier, P., Lodygensky, O., Magniette, F., et~al.: Mpich-v:
  Toward a scalable fault tolerant mpi for volatile nodes. In: SC'02:
  Proceedings of the 2002 ACM/IEEE Conference on Supercomputing. pp. 29--29.
  IEEE (2002)

\bibitem{bosilca2016failure}
Bosilca, G., Bouteiller, A., Guermouche, A., Herault, T., Robert, Y., Sens, P.,
  Dongarra, J.: Failure detection and propagation in hpc systems. In: SC'16:
  Proceedings of the International Conference for High Performance Computing,
  Networking, Storage and Analysis. pp. 312--322 (2016)

\bibitem{doi:10.1177/1094342017711505}
Bosilca, G., Bouteiller, A., Guermouche, A., Herault, T., Robert, Y., Sens, P.,
  Dongarra, J.: A failure detector for hpc platforms. The International Journal
  of High Performance Computing Applications  \textbf{32}(1),  139--158 (2018).
  \doi{10.1177/1094342017711505},
  \url{https://doi.org/10.1177/1094342017711505}

\bibitem{bouteiller2015plan}
Bouteiller, A., Bosilca, G., Dongarra, J.J.: Plan b: Interruption of ongoing
  mpi operations to support failure recovery. In: Proceedings of the 22nd
  European MPI Users' Group Meeting. p.~11 (2015)

\bibitem{cao2016system}
Cao, J., Arya, K., Garg, R., Matott, S., Panda, D.K., Subramoni, H., Vienne,
  J., Cooperman, G.: System-level scalable checkpoint-restart for petascale
  computing. In: 2016 IEEE 22nd International Conference on Parallel and
  Distributed Systems (ICPADS) (2016)

\bibitem{doi:10.1002/cpe.4863}
Chakraborty, S., Laguna, I., Emani, M., Mohror, K., Panda, D.K., Schulz, M.,
  Subramoni, H.: Ereinit: Scalable and efficient fault-tolerance for
  bulk-synchronous mpi applications. Concurrency and Computation: Practice and
  Experience  \textbf{0}(0),  e4863. \doi{10.1002/cpe.4863},
  \url{https://onlinelibrary.wiley.com/doi/abs/10.1002/cpe.4863}, e4863
  cpe.4863

\bibitem{Dongarra:2011:IES:1943326.1943339}
Dongarra, J., Beckman, P., Moore, T., Aerts, P., Aloisio, G., Andre, J.C.,
  Barkai, D., Berthou, J.Y., Boku, T., Braunschweig, B., Cappello, F., Chapman,
  B., Chi, X., Choudhary, A., Dosanjh, S., Dunning, T., Fiore, S., Geist, A.,
  Gropp, B., Harrison, R., Hereld, M., Heroux, M., Hoisie, A., Hotta, K., Jin,
  Z., Ishikawa, Y., Johnson, F., Kale, S., Kenway, R., Keyes, D., Kramer, B.,
  Labarta, J., Lichnewsky, A., Lippert, T., Lucas, B., Maccabe, B., Matsuoka,
  S., Messina, P., Michielse, P., Mohr, B., Mueller, M.S., Nagel, W.E.,
  Nakashima, H., Papka, M.E., Reed, D., Sato, M., Seidel, E., Shalf, J.,
  Skinner, D., Snir, M., Sterling, T., Stevens, R., Streitz, F., Sugar, B.,
  Sumimoto, S., Tang, W., Taylor, J., Thakur, R., Trefethen, A., Valero, M.,
  Van Der~Steen, A., Vetter, J., Williams, P., Wisniewski, R., Yelick, K.: The
  international exascale software project roadmap. Int. J. High Perform.
  Comput. Appl.  \textbf{25}(1),  3--60 (Feb 2011).
  \doi{10.1177/1094342010391989},
  \url{http://dx.doi.org/10.1177/1094342010391989}

\bibitem{Gamell2014Fenix}
Gamell, M., Katz, D.S., Kolla, H., Chen, J., Klasky, S., Parashar, M.:
  Exploring automatic, online failure recovery for scientific applications at
  extreme scales. In: Proceedings of the International Conference for High
  Performance Computing, Networking, Storage and Analysis. pp. 895--906. SC
  '14, IEEE Press, Piscataway, NJ, USA (2014). \doi{10.1109/SC.2014.78},
  \url{https://doi.org/10.1109/SC.2014.78}

\bibitem{gamell2015local}
Gamell, M., Teranishi, K., Heroux, M.A., Mayo, J., Kolla, H., Chen, J.,
  Parashar, M.: Local recovery and failure masking for stencil-based
  applications at extreme scales. In: SC'15: Proceedings of the International
  Conference for High Performance Computing, Networking, Storage and Analysis.
  pp. 1--12 (2015)

\bibitem{hargrove2006berkeley}
Hargrove, P.H., Duell, J.C.: Berkeley lab checkpoint/restart (blcr) for linux
  clusters. In: Journal of Physics: Conference Series. vol.~46, p.~494 (2006)

\bibitem{herault2015practical}
Herault, T., Bouteiller, A., Bosilca, G., Gamell, M., Teranishi, K., Parashar,
  M., Dongarra, J.: Practical scalable consensus for pseudo-synchronous
  distributed systems. In: SC'15: Proceedings of the International Conference
  for High Performance Computing, Networking, Storage and Analysis. pp. 1--12
  (2015)

\bibitem{hori2015sliding}
Hori, A., Yoshinaga, K., Herault, T., Bouteiller, A., Bosilca, G., Ishikawa,
  Y.: Sliding substitution of failed nodes. In: Proceedings of the 22nd
  European MPI Users' Group Meeting. p.~14. ACM (2015)

\bibitem{katti2015scalable}
Katti, A., Di~Fatta, G., Naughton, T., Engelmann, C.: Scalable and fault
  tolerant failure detection and consensus. In: Proceedings of the 22nd
  European MPI Users' Group Meeting. p.~13 (2015)

\bibitem{katti2018epidemic}
Katti, A., Di~Fatta, G., Naughton, T., Engelmann, C.: Epidemic failure
  detection and consensus for extreme parallelism. The International Journal of
  High Performance Computing Applications  \textbf{32}(5),  729--743 (2018)

\bibitem{kohl2019scalable}
Kohl, N., H{\"o}tzer, J., Schornbaum, F., Bauer, M., Godenschwager, C.,
  K{\"o}stler, H., Nestler, B., R{\"u}de, U.: A scalable and extensible
  checkpointing scheme for massively parallel simulations. The International
  Journal of High Performance Computing Applications  \textbf{33}(4),  571--589
  (2019)

\bibitem{Laguna:2014}
Laguna, I., Richards, D.F., Gamblin, T., Schulz, M., de~Supinski, B.R.:
  Evaluating user-level fault tolerance for mpi applications. In: Proceedings
  of the 21st European MPI Users' Group Meeting. pp. 57:57--57:62. EuroMPI/ASIA
  '14, ACM, New York, NY, USA (2014). \doi{10.1145/2642769.2642775},
  \url{http://doi.acm.org/10.1145/2642769.2642775}

\bibitem{Laguna:2014:EUF:2642769.2642775}
Laguna, I., Richards, D.F., Gamblin, T., Schulz, M., de~Supinski, B.R.:
  Evaluating user-level fault tolerance for mpi applications. In: Proceedings
  of the 21st European MPI Users' Group Meeting. pp. 57:57--57:62. EuroMPI/ASIA
  '14, ACM, New York, NY, USA (2014). \doi{10.1145/2642769.2642775},
  \url{http://doi.acm.org/10.1145/2642769.2642775}

\bibitem{laguna2016evaluating}
Laguna, I., Richards, D.F., Gamblin, T., Schulz, M., de~Supinski, B.R., Mohror,
  K., Pritchard, H.: Evaluating and extending user-level fault tolerance in mpi
  applications. The International Journal of High Performance Computing
  Applications  \textbf{30}(3),  305--319 (2016)

\bibitem{doi:10.1177/1094342015623623}
Laguna, I., Richards, D.F., Gamblin, T., Schulz, M., de~Supinski, B.R., Mohror,
  K., Pritchard, H.: Evaluating and extending user-level fault tolerance in mpi
  applications. The International Journal of High Performance Computing
  Applications  \textbf{30}(3),  305--319 (2016).
  \doi{10.1177/1094342015623623},
  \url{https://doi.org/10.1177/1094342015623623}

\bibitem{losada2017resilient}
Losada, N., Cores, I., Mart{\'\i}n, M.J., Gonz{\'a}lez, P.: Resilient mpi
  applications using an application-level checkpointing framework and ulfm. The
  Journal of Supercomputing  \textbf{73}(1) (2017)

\bibitem{6903615}
{Martino}, C.D., {Kalbarczyk}, Z., {Iyer}, R.K., {Baccanico}, F., {Fullop}, J.,
  {Kramer}, W.: Lessons learned from the analysis of system failures at
  petascale: The case of blue waters. In: 2014 44th Annual IEEE/IFIP
  International Conference on Dependable Systems and Networks. pp. 610--621
  (June 2014). \doi{10.1109/DSN.2014.62}

\bibitem{6494566}
{Mohror}, K., {Moody}, A., {Bronevetsky}, G., {de Supinski}, B.R.: Detailed
  modeling and evaluation of a scalable multilevel checkpointing system. IEEE
  Transactions on Parallel and Distributed Systems  \textbf{25}(9),  2255--2263
  (Sep 2014). \doi{10.1109/TPDS.2013.100}

\bibitem{pauli2014fault}
Pauli, S., Kohler, M., Arbenz, P.: A fault tolerant implementation of
  multi-level monte carlo methods. Parallel computing: Accelerating
  computational science and engineering (CSE)  \textbf{25},  471--480 (2014)

\bibitem{sankaran2005lam}
Sankaran, S., Squyres, J.M., Barrett, B., Sahay, V., Lumsdaine, A., Duell, J.,
  Hargrove, P., Roman, E.: The lam/mpi checkpoint/restart framework:
  System-initiated checkpointing. JHPCA  \textbf{19}(4),  479--493 (2005)

\bibitem{shahzad2018craft}
Shahzad, F., Thies, J., Kreutzer, M., Zeiser, T., Hager, G., Wellein, G.:
  Craft: A library for easier application-level checkpoint/restart and
  automatic fault tolerance. IEEE Transactions on Parallel and Distributed
  Systems  \textbf{30}(3),  501--514 (2018)

\bibitem{subasi2018unified}
Subasi, O., Martsinkevich, T., Zyulkyarov, F., Unsal, O., Labarta, J.,
  Cappello, F.: Unified fault-tolerance framework for hybrid task-parallel
  message-passing applications. The International Journal of High Performance
  Computing Applications  \textbf{32}(5),  641--657 (2018)

\bibitem{SULTANA20191}
Sultana, N., Rüfenacht, M., Skjellum, A., Laguna, I., Mohror, K.: Failure
  recovery for bulk synchronous applications with mpi stages. Parallel
  Computing  \textbf{84},  1 -- 14 (2019).
  \doi{https://doi.org/10.1016/j.parco.2019.02.007},
  \url{http://www.sciencedirect.com/science/article/pii/S0167819118303260}

\bibitem{teranishi2014toward}
Teranishi, K., Heroux, M.A.: Toward local failure local recovery resilience
  model using mpi-ulfm. In: Proceedings of the 21st european mpi users' group
  meeting. p.~51 (2014)

\bibitem{wang2018fault}
Wang, Z., Gao, L., Gu, Y., Bao, Y., Yu, G.: A fault-tolerant framework for
  asynchronous iterative computations in cloud environments. IEEE Transactions
  on Parallel and Distributed Systems  \textbf{29}(8),  1678--1692 (2018)

\bibitem{6264677}
{Zheng}, G., {Xiang Ni}, {Kalé}, L.V.: A scalable double in-memory checkpoint
  and restart scheme towards exascale. In: IEEE/IFIP International Conference
  on Dependable Systems and Networks Workshops (DSN 2012). pp.~1--6 (June
  2012). \doi{10.1109/DSNW.2012.6264677}

\bibitem{Zheng:2006:PEA:1131322.1131340}
Zheng, G., Huang, C., Kal{\'e}, L.V.: Performance evaluation of automatic
  checkpoint-based fault tolerance for ampi and charm++. SIGOPS Oper. Syst.
  Rev.  \textbf{40}(2),  90--99 (Apr 2006). \doi{10.1145/1131322.1131340},
  \url{http://doi.acm.org/10.1145/1131322.1131340}

\end{thebibliography}


\end{document}